\documentclass[doublespacing]{elsart}
\usepackage[english]{babel}

\usepackage{epsfig}
\usepackage{graphics}

\usepackage{amssymb}

\begin{document}

\begin{frontmatter}

\title{The phenomena of spin rotation and depolarization of high-energy particles in bent
and  straight crystals at Hadron Collider (LHC) and Future
Circular Collider (FCC) energies and the possibility to measure
the anomalous magnetic moments of short-lived particles (charm and
beauty baryons)}

\author{V.G. Baryshevsky}

\address{Research Institute for Nuclear Problems, Belarusian State
University, 11~Bobruiskaya Str., Minsk 220030, Belarus}
\ead{bar@inp.bsu.by, v\_baryshevsky@yahoo.com}

\begin{abstract}
We study the phenomena of spin rotation and depolarization of
high-energy particles in  crystals  in the range of high energies
that will be available at Hadron Collider (LHC) and Future
Circular Collider (FCC). It is shown that these phenomena can be
used to measure the anomalous magnetic moments of short-lived
particles in this range of energies.  We also demonstrate that the
phenomenon of particle spin depolarization in crystals provides a
unique possibility of measuring the anomalous magnetic moment of
negatively-charged particles (e.g., beauty baryons), for which the
channeling effect is hampered due to far more rapid dechanneling
as compared to that for  positively-charged particles. Channeling
of particles in either straight or bent crystals with polarized
nuclei could be used for polarization and the analysis thereof of
high-energy particles.
\end{abstract}

\end{frontmatter}

\section{Introduction}

 The magnetic moment is an important characteristic of elementary
 particles, but the magnetic moments of many particles (e.g., charm and beauty baryons) are, as
 yet, not measured. This is because for  particles with a  short lifetime $\tau $
 ($\tau =2\cdot 10^{-13}$\,s for $\Lambda^+_c$ and $\tau= 3.5\cdot 10^{-13}$\,s
 for  $\Xi^+_c$),  the decay length $l\sim 3\div 4$ cm  if
 the energy acquired through the production reaction
 equals 1 TeV \cite{rins_104}. For this reason,  the anomalous magnetic moments of
 short-lived particles cannot be  measured with conventional
 methods.

  The existence of the spin rotation phenomenon for high-energy
  particles moving in bent crystals in the channeling regime was first established in \cite{chan_10}.
    The
  spin rotation angle is determined by the anomalous magnetic
  moment $\mu'$ and, according to \cite{chan_10},  can attain values  as high as
  many radians. The unique feature of the effect is that for
  Lorentz-factors $\gamma\gg 1$, the  value of the limit
  spin rotation angle per unit length of 1 cm in a crystal with a limit radius of
  curvature $R_{cr}$ at which the particle still  moves in the channeling
  regime without being expelled from the channel by the
  centrifugal force is independent of particle energy.

  The idea advanced  in \cite{chan_10} was experimentally verified  and confirmed for $\Sigma^+$ hyperons in
  Fermilab \cite{nim06_n4,nim06_n4a,nim06_n4b}. A detailed analysis of the experiment \cite{nim06_n4,nim06_n4a,nim06_n4b}
  given in \cite{rins_104,rins_105,baublis} suggests the feasibility of using the
  spin rotation effect in bent crystals to measure the
  anomalous magnetic moment of short-lived charm baryons $\Lambda^+_c$
  and $\Xi^+_c$.
In  \cite{rins_104}, a thorough consideration was given to  charm
baryons $\Lambda^+_c$ produced by a beam
 of  protons whose  momentum in a tungsten
 target was $800$\,GeV/s. The characteristic momentum of the produced $\Lambda^+_c$ was $200\div
 300$ \, GeV/s (characteristic Lorentz factor $\gamma\simeq 153$).
 At the same time, it was stated in \cite{baublis} that since the beauty quark has a negative charge,
 the beauty baryons have a negative charge, too, and cannot be
 channeled.
 As a result, the spin rotation  phenomena for them
 does  not exists. Positively-charged antiparticles could be
 channeled, but the production rates at LHC
 energies would be too small. It is noteworthy here that with growing particle
 energy the dechanneling length is increased not only for
 positively-charged particles, but also for their negatively charged counterparts.
 What is more, successful experiments
 recently demonstrated the possibility to deflect negatively-charged particles at GeV energies \cite{tikhomirov}.
 For this reason, in light of the projected increase in energy at LHC and
 FCC,
it seems pertinent to return to  the issue of using the spin
rotation effect in bent crystals for measuring the magnetic moment
of not only positively- but also negatively-charged particles.

The effect of spin depolarization of high-energy particles
scattered  by crystal axes (planes)  provides another possibility
of measuring the anomalous magnetic moment of short-lived,
high-energy particles \cite{hyperon_10}. What is more,  an
important advantage of the spin depolarization effect is that it
applies to both positively- and negatively-charged particles.
Thus, this effect enables magnetic moment measurements of not only
 positively-charged baryons, but also  beauty baryons, whose
negative charge and rapid  dechanneling make such measurements
quite a challenge   even in the range of high energies.

This paper considers the effects of  particle spin rotation and
spin depolarization in crystals at LHC and FCC energies. It is
shown that these effects can be used for measuring anomalous
magnetic moments of short-lived particles in this range of
energies.
It is also demonstrated that  the effect of spin depolarization of
high-energy particles scattered  by crystal axes (planes)  can be
used effectively for  measuring the anomalous magnetic moments of
negatively-charged beauty baryons, too.
Because the depolarization effect is possible at wider angles as
compared with the angles of capture into channeling regime
(determined by the Lindhard angle),  the number of usable
 events in this case appears to be greater than in the
experiments with bent crystals, thus giving us  hope for measuring
the anomalous magnetic moment of beauty baryons.

Before proceeding to the consideration of the possibility to use
the phenomena of spin rotation and depolarization of high-energy
particles moving in bent or straight crystals for measuring the
anomalous magnetic moment of charm and beauty baryons, let us
recall briefly the theory underlying these phenomena, as it is
stated in \cite{chan_10,hyperon_10,NO}.

\section{Spin rotation of relativistic particles passing through a
bent crystal and measurement of the magnetic moment of short-lived
particles}

In 1976, E. Tsyganov \cite{chan_8a,chan_8b} demonstrated that
high-energy particles may move in the channeling regime in a bent
crystal, thus moving along the curved path and deviating from the
initial direction. In 1979, the predicted effect was
experimentally verified in JINR (Dubna, Russia) \cite{rins_108}:
the proton beam with the kinetic energy of 8.4\,GeV was rotated
through the angle of 26\,mrad by means of the silicon crystal of
length 2.0\,cm. At present, the experiments on channeling in bent
crystals are performed at the world's largest accelerators. In
these experiments a great variety of crystal optical elements are
widely used to manipulate the beams of high-energy particles.

\begin{figure}[htbp]
\centering \epsfxsize = 6 cm
\centerline{\epsfbox{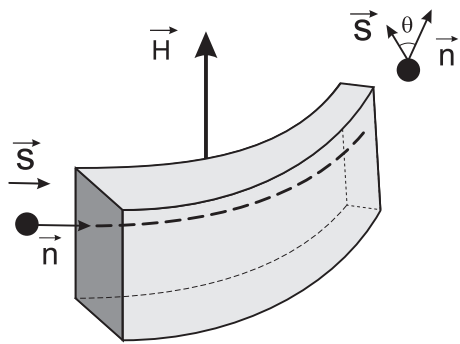}} \caption{Spin rotation in
bent crystal } \label{Fig16.1}
\end{figure}

The next step forward  was  made in 1979 by V. Baryshevsky, who
demonstrated  \cite{chan_10} that the spin of a particle moving in
a bent crystal rotates with respect to the momentum direction
Figure (\ref{Fig16.1}).

Indeed, in a bent crystal a particle moves along a curved path
under the action of the electric field $\vec{E}$ induced by the
crystallographic plane. In the instantaneous rest frame of the
particle, due to relativistic effects, this field produces a
magnetic field $\vec{H}$, which acts on the particle magnetic
moment, thus causing spin rotation in this field (Figure
\ref{Fig16.1}).

Owing to an extremely large magnitude of the field $\vec{E}\approx
10^{7}\div 10^{9}$ CGSE  and, as a result, a large magnitude of
the field $\vec{H}\approx 10^{7}\div 10^{9}$ Gauss, the spin
rotation angle in the crystal of several centimeters in length can
reach several radians.

Now, let us proceed to a more detailed consideration of the effect
of particle spin rotation in a bent crystal.

From the  Bargmann-Michel-Telegdi (BMT) equation \cite{landau}
follows that as the energy of particles increases, their spin
precession frequency in the external magnetic field changes, being
determined in the ultrarelativistic case by the anomalous magnetic
moment. As a result, for example, in a magnetic field
$H=10^{4}$\,Gs, the electron (proton) spin precession frequency
$\omega=2\mu^{\prime}H/\hbar=10^{8}$\,s$^{-1}$ ($\mu^{\prime}$ is
the anomalous part of the magnetic moment), and the spin rotation
angle per one centimeter path length $l$ is just
\[
\vartheta=\omega\frac{l}{c}\approx 10^{-2}\,\mbox{rad}\,.
\]
As mentioned above, a relativistic particle in a bent crystal
"senses" a much greater magnetic field. As the particle energy is
increased (particle wavelength is decreased), the quasi-classical
character of particle motion enables one to apply the laws of
motion determined by classical mechanics.

If a crystal is nonmagnetic, the BMT equation for the spin
polarization vector $\vec{\zeta}=\vec{s}/s$ can be written in the
form \cite{chan_10,hyperon_10,NO}
\begin{equation}
\label{chan_25.1}
\frac{d\vec{\zeta}}{dt}=\frac{2\mu^{\prime}}{\hbar}[\vec{E}(\vec{\zeta}\vec{n})-\vec{n}(\vec{\zeta}\vec{E})]\,,
\end{equation}
where $\vec{E}$ is the electric field at the point of particle
location and $\vec{n}=\vec{v}/c$ (where $\vec{v}$ is the particle
velocity).
Let us recall here that vector $\vec \zeta$ characterizes the
polarization of a particle in its "instantaneous" rest frame.

 Intracrystalline
fields $\vec{E}$ are large, reaching the values of $10^{7}$ CGSE
and even greater. Therefore  from (\ref{chan_25.1}) follows  that
for constant intracrystalline fields, the spin precession
frequency could reach $10^{11}$\,s$^{-1}$ and the angle
$\vartheta$ could be of the order of $10$\,rad/\,cm.

However, when a particle moves through a crystal in arbitrary
direction, the field $\vec{E}$, in a similar manner as in the
amorphous medium, takes on random values at the particle location
point. As a consequence, such a field causes spin depolarization.

Under channeling conditions, the situation is basically different.
If the crystal bending radius is  $\rho_{0}$, then, for example, a
beam of high--energy protons will change its direction following
the crystal bend \cite{chan_8a,chan_8b}, i.e., the particle will
move along a curved path. The stated motion is due to a constant
mean electric field acting on a particle in a bent crystal.

Equation (\ref{chan_25.1}) for a particle moving in a crystal, for
example, in a planar channel bent to a radius of curvature
$\rho_{0}$ around the $y$-axis, has a form ($v_{y}=0$, $E_{y}=0,$
the trajectory lies in the $x$, $z$ plane)
\begin{equation}
\label{chan_25.2}
\frac{d\zeta_{x(z)}}{dt}=\pm\frac{2\mu^{\prime}}{\hbar}(E_{x}n_{z}-E_{z}n_{x})\zeta_{x(z)}\,.
\end{equation}
The position vector $\vec{\rho}=(x,z)$ of a particle in such a
channel rotates about the $y$-axis with the frequency
$\Omega=c/\rho_{0}$. Its magnitude oscillates about the particle
equilibrium position $\rho_{0}^{\prime}$ in the channel  with the
frequency $\Omega_{k}$, the  amplitude $a$, and the initial phase
$\delta$. In the explicit form $x=\rho(t)\cos\Omega t$,
$z=\rho(t)\sin\Omega t$, and $\rho(t) = \rho_{0}^{\prime}+ a
\cos(\Omega_{k}t+\delta)$.
We point out that, due to the presence of centrifugal forces in a
bent crystal, the equilibrium point $\rho_{0}^{\prime}$ does not
coincide with the position $\rho_{0}$ of the minimum of the
electrostatic  potential $\varphi(\rho)$ of the channel. For
example,  for a particle moving in a harmonic well
\[
\varphi=-\frac{k}{e}\frac{(\rho-\rho_{0})^{2}}{2}\,,
\]
 where  the constant $k$ is determined by the crystal properties.

Integration of (\ref{chan_25.2}) in the polar coordinate system
gives ($|\vec{\zeta}|=1$, $\vec{E}=-\vec{\nabla}\varphi$):
\begin{equation}
\label{chan_25.3} \zeta_{z(x)}=\begin{array}{c}
               \cos \\
               \sin
             \end{array}
\left\{\frac{2\mu^{\prime}\Omega}{\hbar
c}\int^{t}_{0}\rho\frac{d\varphi}{d\rho}dt^{\prime}+\arctan\frac{\zeta_{x}(0)}{\zeta_{z}(0)}\right\}\,.
\end{equation}
For a harmonic well  we can write (\ref{chan_25.3}) accurate up to
the terms of the order $(\rho_{0}^{\prime}-\rho_{0})/\rho_{0}$ and
$a\,\rho_{0}^{-1}\ll 1$  in the form
\begin{equation}
\label{chan_25.4} \zeta_{z(x)}(t)=\begin{array}{c}
               \cos \\
               \sin
             \end{array}
\left\{\omega
t+\beta[\sin(\Omega_{k}t+\delta)-\sin\delta]+\arctan\frac{\zeta_{x}(0)}{\zeta_{z}(0)}\right\}\,,
\end{equation}
where
\begin{equation}
\label{5n} \omega=\frac{2\mu^{\prime}}{\hbar}E(\rho_{0}^{\prime})
\end{equation}
 and
\begin{equation}
\label{6n}
E(\rho_{0}^{\prime})=-\frac{k}{e}(\rho_{0}^{\prime}-\rho_{0})
 \end{equation}
 is the electric field at the location point of the particle center of equilibrium in the bent crystal;
 \begin{equation}
 \label{7n}
 \beta=-\frac{2\mu^{\prime}k a}{\hbar e\Omega_{k}}\,.
 \end{equation}

The coefficient $\beta$ in (\ref{chan_25.4}) is small (for Si, the
coefficient $k=4\cdot 10^{17}$\,eV/\,cm$^{2}$, $\Omega_{k}\simeq
10^{13}$\,s$^{-1}$ for protons with the energy  $W\sim 100$\,GeV,
as a result, $\beta=\simeq 10^{-2}$). Neglecting the term
containing $\beta$, we obtain that the spin rotates with frequency
$\omega$ (with growing energy $\Omega_{k}\sim 1/\sqrt{W}$, the
coefficient $\beta\sim \sqrt{W}$ increases, and the spin rotation
turns into oscillations at frequency $\omega$ and  the frequencies
multiple of $\Omega_{k}$).  Due to a large magnitude of the field
$E(\rho_{0}^{\prime})$ curving the particle trajectory
($E(\rho_{0}^{\prime})\sim 10^{7}-10^{9}$ CGSE), the frequency
$\omega\simeq 10^{11}-10^{13}$\,s$^{-1}$, and the rotation angle
$\vartheta\sim 10-10^{3}$\,rad/\,cm.

If the radius of curvature $\rho_{0}\rightarrow \infty$ (a
straight channel), then only spin oscillations due to the term
containing $\beta$ remain. In this case, a significant spin
rotation   occurs only at high energies: for $100$\,TeV protons,
for example, $\beta \simeq 0.3$.

Thus, according to (\ref{chan_25.1}), the spin precession
frequency of a particle moving in a bent crystal is
\begin{equation}
\label{8n}
 \omega=\frac{2\mu' E}{\hbar}.
\end{equation}

As stated above, this expression for the frequency follows
immediately if we consider what field acts on the spin in the
particle's instantaneous rest frame. As a result of relativistic
transformations, the electric field that is transverse relative to
the particle velocity in the particle rest frame  generates a
magnetic field  $H=\gamma E$ that is orthogonal to $E$ and has a
magnitude equal to $\gamma E$. The spin precession frequency
associated with  the anomalous magnetic moment in the particle
rest frame is
\begin{equation}
\label{9n}
 \omega' = \frac{2\mu' H}{\hbar}=\frac{2\mu'\gamma
E}{\hbar}.
\end{equation}

The spin precession frequency in the laboratory frame  is
\begin{equation}
\label{10n}
 \omega = \frac{\omega'}{\gamma}=\frac{2\mu' E}{\hbar}.
\end{equation}

It is noteworthy that with zero particle's anomalous magnetic
moment $\mu'$,  the particle spin precession frequency would equal
the orbital rotation  frequency of the momentum, and the particle
spin direction would follow that of the momentum. If $\mu'\neq 0$,
the angle between the polarization and momentum direction vectors
will change.
The angle $\vartheta_s$ of particle spin rotation relative to the
particle momentum direction  is $\vartheta_s= \omega T=\omega
\frac{L}{c}$, where the time  $T=\frac{L}{c}$, with $L$ being the
path length traveled by the channeled particle in the bent
crystal. From this follows that the rotation angle per unit path
length
\begin{equation}
\label{11n}
 \vartheta_{s1}= \frac{\omega}{c} = \frac{2\mu'
E}{\hbar c}.
\end{equation}

Let us also remember that the  magnetic moment of  the particle
with spin $S$ is related to the gyromagnetic (Land\'{e}) factor
$g$ (see, e.g., \cite{landau}) as
\begin{equation}
 \label{12n}
 \mu =
\frac{e\hbar}{2mc} g S,
\end{equation}
where $m$ is the particle mass.
When $S=\frac{1}{2}$ we can write
 \begin{equation}
 \label{13n}
 \mu = \frac{e\hbar}{2 m c}+\frac{e\hbar}{2 m c}\frac{g-2}{2}=\mu
 _{B} +\mu' ,\qquad  \mu' = \frac{g-2}{2}\mu_B,
\end{equation}
where $\mu_B$ is the Bohr magneton. Consequently, we have
\begin{equation}
\label{e1} \vartheta_s=\frac{2\mu' E}{\hbar c}L = \frac{2 e
\hbar}{\hbar c\cdot 2mc}\,\,\frac{g-2}{2} E L= \frac{g
-2}{2}\frac{e E L}{mc^2}.
\end{equation}
Here $eE$ is the force responsible for rotating the particle
momentum that should be equal to the centrifugal force $f_c =
\frac{m\gamma c^2}{R}$; $e E = \frac{m\gamma c ^2}{R}$, where $R$
is the radius of curvature  of the channel. From this we have
\begin{equation}
\label{e2} \vartheta_s= \frac{g-2}{2}\,\,\frac{m \gamma c^2} {m
c^2 R}L=\frac{g-2}{2}\gamma \frac{L}{R}
\end{equation}
We shall take into account that $\frac{L}{R}$ equals the value of
the momentum's angle of rotation $\vartheta_p$. Hence,
\begin{equation}
\label{e16} \vartheta_s = \frac{g-2}{2}\gamma \vartheta_p.
\end{equation}

Equation (\ref{e16}) was derived by V. Lyuboshitz using the BMT
equation \cite{Lyuboshitz}.  It follows from (\ref{e2}) that the
spin rotation angle per unit path length (the angle of rotation
per 1 cm) is
\begin{equation}
\label{e3} \vartheta_{s1}= \frac{g-2}{2} \frac{\gamma}{R}
\end{equation}
On the other hand,  the equality $e E =\frac{mc^2\gamma}{R}$
yields $R= \frac{mc^2\gamma}{e E}$.

The quantity $|eE|= |e\frac{d\varphi}{d\rho}|= u'$, where $u $ is
the particle's potential energy in the channel. As a consequence,
we have
\[
R =\frac{mc^2}{u'}\gamma.
\]
The minimum value of the radius of curvature (the critical radius)
$R_{cr}$  required for the motion in a bent channel to be still
possible for ceratin $\gamma$ is achieved at maximum $u'_{max}$.
Consequently, the maximum rotation angle per unit \textbf{length}
is
\[
\vartheta_{s1}^{max}= \frac{g-2}{2}\frac {\gamma}{R_{cr}}= \frac
{g-2}{2}\frac{u'_{max}}{mc^2}.
\]

As is seen, the maximum spin rotation angle $\vartheta_{s1}^{max}$
is energy-independent. However, at $R=R_{max}$, the capturing
acceptance (the fraction of particles captured into channeling
regime) vanishes \cite{biryukov}.
For certain $R>R_{cr}$, the spin rotation angle can be written in
the form
\begin{equation}
\label{e4} \vartheta_{s1}=
\frac{g-2}{2}\frac{\gamma}{R_{cr}}\frac{R_{cr}}{R}
=\vartheta_{s1}^{max}\frac{R_{cr}}{R}=\frac{g-2}{2}\frac{u'_{max}}{mc^2}\frac{R_{cr}}{R}.
\end{equation}
As a result, we can see that for the given ratio $\frac{R_cr}{R}$,
the angle $\vartheta_{s1}$ is energy-independent, too.
The dechanneling length increases as the energy is increased
\cite{biryukov}. This enables us to increase the rotation angle by
increasing the length of the bent crystal channel
\[
\vartheta_s=\vartheta_{s1} L,\qquad L\leq L_D,
\]
where $L_D$ is the dechanneling length. For protons with energies
higher than $150\div 200$ GeV, the length $L_D > 10 $ cm.
According to \cite{biryukov}, channeling of protons in 15-cm-long
Si crystals is currently available. As reported in
\cite{biryukov}, for Si,  $u'= 5$ GeV cm$^{-1}$, and hence the
rotation angle per 1 cm path length in a Si crystal is
\[
\vartheta_{s1}= \frac{g-2}{2}\,\,\,\frac{5 \,\,\mathrm{GeV\,
cm}^{-1}}{mc^2} \frac{R_{cr}}{R}.
\]
According to the estimates \cite{rins_104}, for a particle with
mass $\Lambda^+_c\simeq 2.29$ GeV, the $g$-factor lies in the
range $1.36\div 2.45$. From this we get the minimum value of the
ratio $\frac{g-2}{2}\simeq 0.32$. Taking
$\frac{R_c}{R}=\frac{1}{3} $ (in which case the dechanneling
length is maximum \cite{biryukov}), we obtain that for
$\Lambda^+_c$, the rotation angle per unit length in a bent
silicon crystal equals
\[
\vartheta_{s1}\simeq 0.46\quad \mathrm{rad}.
\]
For $\frac{R_c}{R}=\frac{1}{10}$, we have $\vartheta_{s1}\simeq
0.1$\,rad; for $L=10$\,cm, we have $\vartheta_{s1}\simeq 1$\,rad.

Thus, if the ratio $\frac{R_c}{R}$ remains constant, the value of
 the spin rotation angle appears to be constant and rather large in a
wide range of energies.
 The decrease in the Lindhard angle $\vartheta_L\sim \frac{1}{\sqrt{\gamma}}$ with increasing  energy  seemingly impairs
 the observation conditions. However, as the energy is increased, the angular
 width $\delta\vartheta$
 in
 which the produced particles move is
 $\delta\vartheta\sim\frac{1}{\gamma}$, and hence  the fraction of
 particles captured into the channel is increased with energy as
 $\frac{\vartheta_L}{\delta\vartheta}\sim \sqrt{\gamma}$. According to \cite{appel,russ}, as the
 energy is increased, the probability of $\Lambda^+_c$ production also
 increases. For this reason, the increase in energy leads to an
 increasing number of usable events, i.e., the number of deflected
 $\Lambda^+_c$ entering the detector. For example, if
 the energy of  $\Lambda^+_c$ is increased from 300 GeV  to 3
 TeV,  the number of produced
 $\Lambda^+_c$ increases tenfold if the linear growth of the
 production cross section for $\Lambda^+_c$, suggested in \cite{appel,russ},  continues  as the energy of the incident protons rises.
 Moreover, the amount  of $\Lambda^+_c$ captured into channeling
 regime is increased threefold.
 As a result, the time required to observe the effect is reduced.


\section{Spin depolarization of relativistic particles traveling
through a crystal}

  When a particle travels through a crystal, the  field $\vec E$
   at the point of particle location  takes on random values. This is due to fluctuations of the position of the nucleus (electron) in
   the   crystal (in particular, under thermal vibrations) and to
  fluctuations of the electric field that come from particle  scattering by various atomic chains. The fluctuations
  of the field $\vec{E}$ lead to the stochastic spin precession frequency in the electric field.

 Owing to the presence of fluctuations of the spin rotation frequency,
 the spin rotation angle of any particle that  has traveled through a crystal
 takes on different values.
  As a result,  after passing through a target,
 every particle has a different direction of the polarization vector $\vec{\zeta}$, and the distribution of the particle spin direction
 in the beam may even become isotropic, i.e., the beam polarization vector goes to zero.
  The stochasticity of $\omega$ leads to spin depolarization  of particles moving in the crystal \cite{hyperon_10}.

 According to
\cite{hyperon_10}, depolarization of particles in the case of
scattering by crystal planes (axes) is determined by the intensity
of multiple scattering by crystal planes (axes).

It was shown that depolarization has practically no effect on the
possibility of measuring the magnetic moment of particles moving
in  bent crystals in channeling regime.

A much stronger depolarization  is observed for the particles that
are not involved in the channeling regime. These particles may
undergo collisions with atomic chains as a whole. This leads to a
sharp increase in the root-mean-square angle of multiple
scattering and, as a consequence, to a large depolarization.

 For example, according to \cite{hyperon_10},
the longitudinal polarization $\langle\zeta_{z}\rangle$ of
particles  undergoing scattering by the axes decreases as
\begin{equation}
\label{nim90_1.39}
\langle\zeta_{z}(l)\rangle=\langle\zeta_{z}(0)\rangle e^{-Gl}\,,
\end{equation}
\begin{equation}
\label{ins_nim90_1.40}
G=\frac{1}{2}\left(\frac{g-2}{2}+\frac{1}{\gamma+1}\right)^{2}\gamma^2\langle\theta^2_1\rangle\,,
\end{equation}
where $\langle\theta^2_1\rangle$ is the mean square angle of
multiple scattering of the particle by the crystal axes per 1
centimeter path length  in the crystal. According to
\cite{synch_7,synch_72}, the mean square angle of particle
scattering by the crystal axes for a
 path length $l$ traveled in the crystal can be written as
\begin{equation}
\label{ins1_nim90_1.40} \langle\theta^2(l)\rangle=\frac{16\pi
D^{2}Z^{2}\hbar^{2}\alpha^{2}N\xi}{M^{2}c^{2}\gamma^2}\frac{\psi_{\mathrm{max}}}{\psi}l\,.
\end{equation}
So we have
\begin{equation}
\label{nim90_1.40}
G=\frac{1}{2}\left(\frac{g-2}{2}+\frac{1}{\gamma+1}\right)^{2}\frac{16\pi
D^{2}Z^{2}
\hbar^{2}\alpha^{2}N\xi}{M^{2}c^{2}}\frac{\psi_{\mathrm{max}}}{\psi}\,,
\end{equation}
where $\alpha=1/137$, $Z$ is the atomic number, $N$ is the number
of atoms per cm$^{3}$, $\xi$ is the numerical coefficient equal to
1.16 for the axial potential in the Lindhard model, and $\psi$ is
the angle between the direction of the particle momentum and the
crystallographic axis;
\[
\psi<\psi_{\mathrm{max}}=\frac{r_s}{d_{1}}\simeq 0.1
\sqrt[3]{Z}\,,
\]
where $r_s$ is the screening radius, $d_{1}$ is the distance
between the axes, and the $z$-axis is parallel to the
crystallographic axis;
\[
D=\frac{e}{e_{0}}
\]
is the ratio  between the charge of a particle and that of an
electron.

 For example, for particles with the parameters of the proton
 (i.e., $(g-2)/2\sim 1.79$, $M\sim 938$\,MeV), we have
for W-type crystals the magnitude ${G\simeq
10^{-4}\psi_{\mathrm{max}}/\psi}$ and, according to the value of
$\psi$, it may reach ${G\simeq 10^{-1}\div 1}$, i.e., for crystal
thickness of the order of 1 cm, the depolarization may amount to a
few tens of percent. Equation (\ref{nim90_1.40}) is valid for
$\psi$, which are larger than the Lindhard angle $\psi_{L}=\sqrt{2
U/Mc^{2}\gamma}$ ($U$ is the averaged axial potential).

It follows from  (\ref{nim90_1.39}) and (\ref{ins_nim90_1.40})
that
\begin{equation}
\label{new} \left|\frac{g-2}{2}+\frac{1}{\gamma+1}\right|=
\sqrt{\frac{2}{\gamma^2\langle\theta^2(l)\rangle}\ln\frac{\langle
\zeta_z(0)\rangle}{\langle\zeta_z(l)\rangle}}.
\end{equation}

When the particle energy corresponds to $\gamma\gg 1$, from
(\ref{nim90_1.39}) and (\ref{ins_nim90_1.40}) we may obtain the
following expression  for $(g-2)$ \cite{hyperon_10}:
\begin{equation}
\label{nim90_1.41}
|g-2|=\sqrt{\frac{8}{\gamma^{2}\langle\theta^{2}(l)\rangle}\ln\frac{\langle\zeta_{z}(0)\rangle}{\langle\zeta_{z}(l)\rangle}}\,,
\qquad \frac{g-2}{2}\gg\frac{1}{\gamma}\,
\end{equation}

Let us note that for electrons and positrons, as reported in
\cite{NO,chan_11,nim90_6}, the experiments on spin rotation in
bent crystals provide a unique possibility for studying the
effects of quantum electrodynamics of a strong field, namely, the
dependence of the anomalous magnetic moment, i.e., $(g-2)/2$, on
the crystal electric field intensity and the particle energy. The
same possibility exists in studying the depolarization of
$e^{\pm}$. Thus, with experimentally determined values of
$\langle\theta^2(l)\rangle$ and $\langle\zeta_{z}(l)\rangle$, we
can find the value of ($g-2$), which gives us another possibility
to measure the anomalous magnetic moment of particles.
It is noteworthy that strong fluctuating fields act on neutral
particles moving at a small angle to the crystal axes (planes),
too, leading to the depolarization of their spin, which fact can
be used to measure the magnetic moment of neutral partticles.

We should  pay attention to the fact that   theoretical
considerations and experimental results  both  reveal that
$\Lambda_c^+$ produced through collisions between nonpolarized
nucleons and nonpolarized nucleons (nuclei) are  polarized, and
their polarization vector is oriented orthogonally to the reaction
plane, i.e., the plane where the momenta of the incident nucleon
and the produced baryon lie.
 A similar situation would  occur in
production of  other baryons--- charm and beauty. This is why the
reaction amplitude $f$ includes the term of the form
\begin{equation}
\label{new1} f\sim \vec S \left[\vec p_N \times \vec p _B \right]
\end{equation}
where $\vec S$ is the spin of the produced particle, $\vec p_N$ is
the momentum of the incident particle, and $\vec p_B$ is the
momentum of the produced baryon.
Consequently, the polarization component $\vec\zeta_{\perp}(0)$ of
the produced baryon that is parallel to the momentum $\vec p_B$
equals zero. Depolarization occurs to the polarization vector
component $\vec\zeta_{\perp}(0)= (\zeta_x(0),\zeta_y(0))$ lying in
the plane transverse relative to vector $\vec p_B$.

Let the produced baryon be incident on the crystal axis at a small
angle $ \psi < \psi_{max}=\frac{R}{d_1}$ (according to
\cite{synch_72}, $\frac{R}{d_1}\sim 0.1\sqrt[3]{z}$).
The baryon's velocity component parallel to the axes
$v_{\parallel}=v \left (1-\frac{\psi^2}{2}\right)$, where $v$ is
the particle velocity. The component $\vec v_{0\perp}$ of the
particle's  initial velocity  is orthogonal to the axis, and
 $v_{0\perp}=v\psi$.
 Let us choose the $x$-axis  along vector $\vec
 v_{0\perp}$. Multiple scattering leads to appearing of the
 velocity's component $v_y$ that is orthogonal to the $x$-axis,
 and the changes in $v_y$ are much greater than those in $v_x$
 in the direction of the $x$-axis. As a consequence, multiple
 scattering by the axes is strongly asymmetric in the plane
 perpendicular to the direction of the baryon momentum \cite{synch_72}.

 This means that the depolarization is also asymmetric in this
 plane, i.e., $\zeta_x(l)\neq \zeta_y(l)$. For this reason, the
 maximum depolarization of the produced baryons can be obtained if
 the crystal axes are oriented in such a way that the initial vector of baryon
polarization is oriented in the direction of the $y$-axis.
According to \cite{hyperon_10}, the component $\zeta_y(l)$ of the
polarization vector relates to $l$ as
\begin{equation}
\label{new2} \zeta_y(l)=\langle \zeta_y(0)\rangle
\exp\left[-\frac{1}{8}(g-2)^2 \gamma^2\langle
\theta_y^2(l)\rangle\right].
\end{equation}

Let us recall that in view of the analysis \cite{synch_72}, the
contributions to multiple scattering by the axes come not only
from the averaged potential, but also from particle scattering by
the fluctuations of the potential of the axes that are due to
thermal oscillations of atoms.
As a result, $\langle \theta^2_y\rangle = \langle
\theta^2_{t\,y}\rangle + \langle \theta_{av\, y}^2\rangle$
\cite{synch_72}, where $\langle\theta^2_{t\,y}\rangle$ is the mean
square angle of multiple scattering due to thermal oscillations
and $\langle \theta_{av\, y}^2\rangle$ is the mean square angle of
multiple scattering due to scattering by averaged over thermal
oscillations potential of the axes.  Let us also note that
$\langle \theta^2_y\rangle \sim \frac{1}{\gamma^2}$, that is,
$\gamma^2\langle \theta^2_y\rangle$ is independent of the particle
energy.
It also follows from (\ref{new2}) that
\begin{equation}
\label{new3} |g-2|=\sqrt{\frac{8}{\gamma^2\langle
\theta_y^2(l)\rangle}\,\ln \frac{\zeta_y(0)}{\zeta_y(l)}}.
\end{equation}

In our further estimations we shall consider only the contribution
coming from multiple scattering by the averaged over thermal
oscillations potential of the axes $\langle \theta_{av\,
y}^2\rangle$ and assume that $\langle \theta_{y}^2\rangle \simeq
\langle \theta_{av\, y}^2\rangle$.

As a result, we have the following expression for the damping rate
\begin{equation}
\label{new4} \zeta_y(l)= \langle \zeta_y(l)\rangle e^{-G l},
\end{equation}
where $G$ is defined by (\ref{nim90_1.40}).

Thus, the depolarization rate in this case is similar to that in
(\ref{nim90_1.39}). Now let $\Lambda^+_c$ be incident on the
crystal. For our estimate we use the value of the $g$-factor given
in \cite{rins_104}: $g\simeq 1.36 \div 2.45 $. The minimum value
is estimated as $\frac{g-2}{2} = 0.32$. So we have the following
estimate for $G$ in tungsten (W): $G\simeq 10^{-1}$ (the beam is
assumed to be incident on the axes at an angle of about several
Lindhard angles); then for the length $l = 10$\,cm, for example,
$Gl\sim 1$. Hence, the polarization in this case decreases by a
factor of $e$, i.e., quite appreciably. More precise estimates
require thorough simulations, but even the estimates given here
indicate that this phenomenon can hopefully be used for
negatively-charged beauty baryons.


\section{Spin rotation and oscillations of high-energy
channeled particles in the effective nuclear potential of a
crystal as one of the possibilities to obtain and analyze the
polarization of high-energy particles}

Now, let us note that when moving in the crystal, such particles
as baryons (antibaryons and nuclei) take part not only in
electromagnetic, but also in strong interactions. As a result, the
spatially periodic effective potential energy of interaction
between a particle and a crystal is determined by two
interactions: Coulomb and strong \cite{NO}.

The effective periodic potential $U(\vec{r})$ can be introduced to
describe the passage of a coherent wave in the crystal:
\begin{equation}
U(\vec{r})=\sum_{\vec{\tau}} U(\vec{\tau}) e^{i \vec{\tau}
\vec{r}}\,,
\end{equation}
where $\vec{\tau}$ is the reciprocal lattice vector of the
crystal,
\begin{equation}
U(\vec{\tau})=\frac{1}{V} \sum_{j} U_{j0}(\vec{\tau})
 e^{-W_j (\vec{\tau})} e^{i \vec{\tau} \vec{r}_j}\,.
\end{equation}
Here $V$ is the volume of the crystal elementary cell, $\vec{r}_j$
is the coordinate of the atom (nucleus) of type $j$ in the crystal
elementary cell, and the squared $e^{-W_j (\vec{\tau})}$ equals
the thermal-factor (i.e., the Debye-Waller factor), well-known for
X-ray scattering \cite{nim06_James}:
\begin{equation}
U_{j0}(\vec{\tau}) =-\frac{2 \pi \hbar^2}{M \gamma}
F_j{(\vec{\tau})}\,,
\end{equation}
where $M$ is the mass of the incident particle, $\gamma$ is its
Lorentz-factor, $F_j{(\vec{\tau})}=F_j{(\vec{k}^\prime-\vec{k}
=\vec{\tau})}$ is the amplitude of elastic coherent scattering of
the particle by the atom, $\vec{k}$ is the wave vector of the
incident wave, and  $\vec{k}^\prime$ is the wave vector of the
scattered wave.

It will be recalled that in the case of a crystal, the imaginary
part of the amplitude $F_{j}(0)$ does not contain the contribution
from the total cross section of elastic coherent scattering. The
imaginary part of $F_{j}(0)$ in a crystal is only determined by
the total cross sections of inelastic processes. This occurs
because in a crystal, unlike in amorphous matter, the wave
elastically scattered at a nonzero angle, due to rescattering by
periodically located centers, is involved in formation of a
coherent wave propagating through the crystal.

Elastic coherent scattering of a particle by an atom is caused by
both the  Coulomb interaction of the particle with the atom's
electrons and nucleus and its nuclear interaction with the
nucleus. Therefore, the scattering amplitude can be presented as a
sum of two amplitudes:
\begin{equation}
F_j{(\vec{\tau})}=F_j^{\mathrm{Coul}}{(\vec{\tau})}+F_j^{\mathrm{nuc}}{(\vec{\tau})}\,,
\end{equation}
where $F_j^{\mathrm{Coul}}{(\vec{\tau})}$ is the amplitude of
particle scattering caused by the Coulomb interaction with the
atom (nucleus) (it contains contributions from the Coulomb
interaction of the particle with the atom along with the
spin-orbit interaction with the Coulomb field of the atom
(nucleus));
$F_j^{\mathrm{nuc}}{(\vec{\tau})}$ is the amplitude of elastic
coherent scattering of the particle caused by the nuclear
interaction
(this amplitude contains the terms independent of the incident
particle spin along with the terms depending on spin of both the
incident particle and nucleus, in particular, spin-orbit
interaction). Therefore, $U(\vec{r})$ and $U(\vec{\tau})$ can also
be expressed:
\begin{eqnarray}
\begin{array}{l}
U{(\vec{r})}=U^{\mathrm{Coul}}{(\vec{r})}+U^{\mathrm{nuc}}{(\vec{r})}\,,
\\
U{(\vec{\tau})}=U^{\mathrm{Coul}}{(\vec{\tau})}+U^{\mathrm{nuc}}{(\vec{\tau})}\,.
\end{array}
\end{eqnarray}
Let us suppose that a high energy particle moves in a crystal at a
small angle to the crystallographic planes (axes) close to the
Lindhard angle ${\vartheta_L \sim \sqrt{U/E}}$ (in a relativistic
case ${\vartheta_L \sim \sqrt{2U/E}}$), where $E$ is the energy of
the particle, $U$ is the height of the potential barrier created
by the crystallographic plane (axis).
This motion is determined by the plane (axis) potential $\hat{U}
(\vec{\rho})$, which could be derived from ${U} (\vec{r})$ by
averaging over the distribution of atoms (nuclei) in the crystal
plane (axis).

As a consequence, the potential $\hat{U} (\vec{\rho})$ for a
particle channeled in a plane (or axis) channel or moving over the
barrier at a small angle, close to the Lindhard angle, can be
expressed as a sum \cite{NO,nim06_Vesti}:
\begin{equation}
\hat{U}
(\vec{\rho})=\hat{U}^{\mathrm{Coul}}{(\vec{\rho})}+\hat{U}^{\mathrm{sp.-orb.}}{(\vec{\rho})}+\hat{U}_{\mathrm{eff}}^{\mathrm{nuc}}
{(\vec{\rho})}+ \hat{U}^{\mathrm{mag}}{(\vec{\rho})}\,,
\end{equation}
where $\hat{U}^{\mathrm{Coul}}$ is the potential energy of
particle Coulomb interaction with the crystallographic plane
(axis),
$\hat{U}^{\mathrm{sp.-orb.}}$ is the energy of spin-orbit
interaction with the Coulomb field of the plane (axis) and
spin-orbit nuclear interaction with the effective nuclear field of
the plane (axis),
 and $\hat{U}_{\mathrm{eff}}^{\mathrm{nuc}}$ is the effective
potential energy of nuclear interaction of the incident particle
with the crystallographic plane (axis).
\begin{equation}
\hat{U}_{\mathrm{eff}}^{\mathrm{nuc}}{(\vec{\rho})}=-\frac{2 \pi
\hbar^2}{M \gamma} N{(\vec{\rho})} \hat{f}(0)\,,
\end{equation}
where $N{(\vec{\rho})}$ is the  density of nuclei at point
$\vec{\rho}=(x,y)$ of the crystallographic plane (axis),  the
$z$-axis  is directed along the crystal axis (along the velocity
component parallel to the crystallographic plane in the case of
plane channeling), $\hat{f}(0)$ is the amplitude of elastic
coherent forward scattering, which depends on the incident
particle spin and nucleus polarization,
$\hat{U}^{\mathrm{mag}}{(\vec{\rho})}$ is the energy of magnetic
interaction of the particle with the magnetic field produced by
the electrons (nuclei).

Thus in describing particle motion in crystals, the contribution
of strong interactions
 to the formation of the effective
potential acting on the particle from the crystallographic planes
(axes) should be considered along with  the Coulomb interaction.
 In the case of high energies,  the particle motion in
the potential $\hat{U}(\rho)$ can be described in the
quasi-classical approximation. The spin-evolution equations for a
particle moving in straight and bent crystals in the presence of
the contribution from
$\hat{U}_{\mathrm{eff}}^{\mathrm{nuc}}(\vec{\rho})$ is discussed
in   \cite{NO,rins_109}.

Let us assume now that the crystal nuclei are polarized.
The cross section of high-energy particle scattering  by a nucleus
(the amplitude of zero-angle scattering) in a crystal with
polarized nuclei depends on the particle spin orientation with
respect to the polarization of the target.
Therefore, the absorption coefficient also depends on particle
spin orientation.

Let a particle move in a crystal at a small angle to the
crystallographic plane (axis) close to the Lindhard angle.
In this case the
 coefficient of particle absorption by nuclei can differ from the
 absorption coefficient for the particle moving in the crystal at a large angle to the
crystallographic plane (axis). Moreover, it can appear to be
greater even for positively-charged particles.
 Let us discuss the possibility to use a polarized crystal to obtain a polarized beam of high-energy
 particles and to analyze the polarization state of high-energy particles \cite{NO,nim06_Vesti}.

 Let us  consider a positively-charged, high-energy particle moving close to the
 top of the potential barrier. Therefore, it moves in the range with nuclei
 density higher than that of an amorphous medium.
In this range, the growth is described by
  \[
  \frac{d}{a_0} \approx
 10^2\,,
 \]
  here $d$ is the lattice period, $a_0$
 is the amplitude of the nucleus thermal oscillations.
For the  over-barrier motion  in the vicinity of the
crystallographic axis,  the growth is described by the relation
\[
\left(\frac{d}{a_0}\right)^2 \approx
 10^4\,.
 \]

 After passing  through the polarized crystal of
 thickness $L$, the particles  moving in the region where the crystal nuclei are located  acquire the  polarization degree
that can be described by formula
 \begin{equation}
 P=\frac{e^{-\rho \sigma_{\uparrow \uparrow}L}-e^{-\rho \sigma_{\uparrow \downarrow}L}}
 {e^{-\rho \sigma_{\uparrow \uparrow}L}+e^{-\rho \sigma_{\uparrow
 \downarrow}L}}\,,
 \end{equation}
where $\sigma_{\uparrow \uparrow}$ is the total scattering cross
section for the particle whose spin is parallel to the
polarization vector of  nuclei and $\sigma_{\uparrow \downarrow}$
is the total scattering cross section for the particle whose spin
is anti-parallel to the polarization vector of nuclei.

To avoid misunderstanding, it will be recalled that the
exponential absorption law holds true if the characteristic
scattering angle of a particle is much larger than the angle at
which the observer from the point of target location may see the
input window of the detector, registering the flux of particles
transmitted through the target. In the opposite case all of the
scattered particles enter the detector, and by
$\sigma_{\uparrow\uparrow}(\sigma_{\downarrow\uparrow})$ one
should mean the total cross section of absorption and reactions
that remove the particle from the incident beam (causing
appreciable change in its energy). The density matrix techniques
should be used as a tool in considering the general case.

The  magnitude of the effect is determined by the parameter
${A=\rho(\sigma_{\uparrow \uparrow}-\sigma_{\uparrow
\downarrow})L}$. With $\rho_{\mathrm{av}}$ being the average
density of nuclei in the crystal, $A=\rho_{\mathrm{av}} \cdot 10^2
\Delta \sigma L$ for a plane and $A=\rho_{\mathrm{av}} \cdot 10^4
\Delta \sigma L$ for an axis.

The length $L_1$ corresponding to $A=1$ is  $L_1 = \frac{1}{\rho
\Delta \sigma}$. If we suppose that $\rho_{\mathrm{av}} \approx
10^{22}$ and $\Delta \sigma \sim 10^{-25}$\,cm$^2$, then for a
plane $L_1 \approx 10$\,cm and for an axis $L_1 \approx
10^{-1}$\,cm. As is seen, in this case a beam of polarized
particles can be obtained using rather small-sized polarized
targets.

Let positively-charged particles move parallel to a
crystallographic plane. In this case,  most of them appear close
to the bottom of the potential well and far from the nuclei,
therefore, the absorption coefficient for this fraction of
particles is smaller comparing to the average crystal absorption.
Nevertheless, in this case there is also a fraction of particles
that moves in the region where the nuclei are located.

A fraction of particles moving near the barrier can be
 sorted out  by means of  angular distribution or by using bent crystals. Moving in a bent crystal, the fraction of
 channeled particles that due to Coulomb repulsion is hardly involved in the nuclear
 interaction is deflected from the initial direction of motion. Thus the particles that have undergone
 the interaction with the nuclei will change the beam moving in the initial
 direction. Because the absorption for this fraction of particles
 increases, it becomes possible to achieve quite a significant
 polarization of the initially nonpolarized beam.

 A bent crystal can be used to increase the interaction of a positively-charged
particle with a nucleus, too.
In this case, the centrifugal forces push the channeled beam to
the range with  high density of nuclei.
A detailed analysis of the polarization state of  transient
particles requires a numerical simulation.

In contrast to  positively charged particles,  negatively charged
particles (for example, $\Omega^{-}$ hyperons, antiprotons, and
beauty hyperons), being channeled, move in the region with high
density of nuclei, therefore, even a very thin polarized crystal
($L \approx 10^{-1}$ cm)
  can be an effective polarizer and polarization analyzer.

Moreover, for a negatively-charged channeled particle, the angle
of spin rotation in the nuclear pseudo-magnetic field of the
polarized crystal \cite{nim06_Vesti} can be larger than  that for
amorphous matter \cite{NO,nim06_Vesti}.
This is the reason for the thin polarized crystals to be used as
nuclear-optical elements that can guide polarization of
high-energy particle beams.

Note that because of the presence of the amplitude similar to that
in (\ref{new1}),  the spin-orbit interaction also leads to
particle polarization, as well as to the left--right asymmetry in
scattering of  polarized particles (both charged and neutral).
The density of nuclei near the axis is much higher than the
density of an amorphous target. This makes the induced
polarization (left--right asymmetry) for particles scattered by
the crystal axes noticeably higher, as compared with amorphous
matter, due to interference of the Coulomb and  nuclear
interactions \cite{NO,nim06_Vesti}.

In this context, intensively developing techniques for collimating
and handling the beams of high-energy particles, using the volume
reflection of charged particles from bent crystal planes deserve
attention \cite{rins_67}-\cite{rins_75}. In particular, according
to \cite{rins_72}, multiple volume reflection from different
planes of one bent crystal becomes possible when particles move at
a small angle with respect to the crystal axis. Such multiple
volume reflection enables a  severalfold increase in  the particle
deflection angle inside one crystal. This effect is experimentally
revealed now, and it is noteworthy that it leads to an increasing
probability of particle-nucleus interaction.

Channeling of particles in either straight or bent crystals with
 polarized nuclei could be used to polarize  high-energy particles or analyze their polarization. The beam of nonpolarized
particles extracted from the storage ring could be significantly
polarized by applying the additional polarized bent (straight)
crystal.

\section{Spin oscillation and the possibility of quadrupole moment
measurements of $\Omega^{-}$-hyperons moving in a crystal}

A few more words should be said about  the possibility of
quadrupole moment measurements for $\Omega^{-}$-hyperons moving in
a  crystal \cite{nim06_n6}.
 Theoretical estimates of the quadrupole moment of short-lived particles, for
 instance of an $\Omega^{-}$-hyperon \cite{hyperon_3,hyperon_4,hyperon_5}, based on the nonrelativistic
 quark model depend on the model parameters and can differ significantly.
 Experimental measurement of the quantity under consideration using the convenient method of passing
 through an inhomogeneous electromagnetic field is at present practically unrealizable because of the
 smallness of quadrupole interaction at attainable strengths of the electromagnetic field.

Despite the seeming simplicity, the use of the effect of spin
oscillation of  $\Omega^{-}$-hyperons channeled in crystals is a
challenging task, too. This is, above all, due to
 the sharp increase of multiple scattering by nuclei in the case of negatively-charged particles ($\Omega^{-}$-hyperons).
 However, it should be emphasized that the difficulties can be overcome to a great extent using the effect of spin
 oscillation occurring in a nonchanneled state of motion (not too far above the critical angle) relative to the chosen
 family of crystallographic planes for particles with spin $I>1/2$.

The equations describing the spin dynamics of a particle
possessing  magnetic and quadrupole moments can be derived using
the results obtained in \cite{NO}. In consequence, the
relativistic equation of spin motion in this case has the
following form:
\begin{equation}
\label{hyperon_1.8}
\frac{d\hat{I}_{i}}{dt}=\epsilon_{ijk}(\Omega_{j}\hat{I}_{k}+\frac{1}{3}e\varphi_{jl}\hat{Q}_{kl})\,,
\end{equation}
where $\Omega_{j}$ is a component of the vector
$\vec{\Omega}=[\frac{1}{2}(g-2)\gamma+\gamma/(\gamma+1)]v^{2}\vec{\Omega}_{0}$
and $\vec{\Omega}_{0}=-(e/M)(\gamma/(\gamma^{2}-1))(\vec{E}\times\vec{v})$.

 Let us consider (\ref{hyperon_1.8}) under the condition that the particle moves in a
 nonchanneled state at an angle $\psi>\psi_{L}$ with respect to the chosen family of
 crystallographic planes, where $\psi_{L}$ is the Lindhard angle. Suppose that the coordinate axis
 $x$ is perpendicular to the given crystallographic planes. Taking into account that $d^{2}U/dxdy$, $d^{2}U/dy^{2}\ll d^{2}U/dx^{2}$, we obtain
\begin{equation}
\label{hyperon_1.9a}
\frac{d\hat{I}_{z}}{dt}=\Omega_{x}\hat{I}_{y}-\Omega_{y}\hat{I}_{x}+\frac{\varphi_{xx}eQ}{2I(2I-1)}\hat{I}_{xy}\,,
\end{equation}
\begin{equation}
\label{hyperon_1.9b}
\frac{d\hat{I}_{y}}{dt}=\Omega_{z}\hat{I}_{x}-\Omega_{x}\hat{I}_{z}-\frac{\varphi_{xx}eQ}{2I(2I-1)}\hat{I}_{xx}\,,
\end{equation}
\begin{equation}
\label{hyperon_1.9c}
\frac{d\hat{I}_{x}}{dt}=\Omega_{x}\hat{I}_{z}-\Omega_{z}\hat{I}_{y}\,.
\end{equation}

 Equations (\ref{hyperon_1.8})--(\ref{hyperon_1.9c}) enable one to analyze whether the
 spin oscillation of a particle moving in
 a nonchanneled state can be used for measuring the electric quadrupole moment of the $\Omega^{-}$-hyperon.

In view of (\ref{hyperon_1.9a})--(\ref{hyperon_1.9c}), when a
particle with a quadrupole moment moves in a crystal, not only
spin oscillation and rotation appear, but also the transitions
between tensor $P_{ik}$ and vector $\vec{P}$ polarizations of the
particle.

 As follows from  (\ref{hyperon_1.8})-(\ref{hyperon_1.9c}), under the initial conditions
 $P_{z}(0)=0$ and {${P_{xy}(0)\neq 0}$}, there is a maximal change of the $\Omega^{-}$-hyperon spin projection
 $P_{z}$ due to the quadrupole interaction.

According to the estimates made in \cite{nim06_n6}, in such
experiments for the $\Omega^{-}$-hyperon beam with  Lorentz factor
$\gamma=100$, intensity $N\approx 10^6$ particle/s, and the beam
divergence angle $\theta_{\mathrm{div}}< 0.4$\,mrad, it is
possible to measure the quadrupole moment $Q$ of the
$\Omega^{-}$-hyperon
 on the level $10^{-27}$\,cm$^{2}$ in a tungsten crystal of length $l=20$\,cm.

 It should be noted that the measurement scheme suggested here for $\Omega^{-}$-hyperon quadrupole
 moment measurements  requires neither a high quality crystal nor a monochromatic hyperon beam!  For the
 realization of this measuring procedure it is quite sufficient to have a crystal with a mosaic spread
 ${\chi< 0.4}$\,mrad or a set of crystals arranged with an exactness of not more than $\chi< 0.4$\,mrad
 relative to the chosen family of crystallographic planes. This condition justifies the possibility of creating
 large crystal systems necessary for carrying out this experiment.

\section{Conclusion}
The analysis given in this paper shows  that the phenomena of spin
rotation and depolarization of high-energy particles in crystals
at energies that will be available  at LHC and FCC can be used to
measure the anomalous magnetic moments of short-lived particles at
such  energies. It has also been demonstrated that  for
negatively-charged particles (e.g., beauty baryons), the
phenomenon of  spin depolarization in crystals is a promising tool
allowing the anomalous magnetic moment
  measurements. Moreover it has been noted that the spin
  depolarization effect occurs for  neutral particles incident
  at small angles to crystal axes (planes), and this opens  the
  potential for magnetic moment measurements of such short-lived
  particles.
Channeling of particles in either straight or bent crystals with
polarized nuclei could be used for  polarization or polarization
analysis of high-energy particles, including neutral particles.
 Depolarization and spin rotation of that fraction of neutral
particles which moves in the region of high concentration of
nuclei can be defined by  particle reactions in the crystal: if at
a certain point the particle is  scattered at a large angle, this
means that it has undergone a collision, and hence it was in the
vicinity of the nuclei. At high energies, particles move along a
straight line. Particles moving between the planes do not undergo
collisions, while those moving along the axis or a plane do.


\end{document}